\documentclass[aip,graphicx,amssymb, reprint]{revtex4-1}
\usepackage[fleqn]{amsmath}
\usepackage{graphicx}
\usepackage{dcolumn}
\usepackage{bm}
\usepackage[colorlinks=true,linkcolor=blue,citecolor=blue,urlcolor=blue]{hyperref}
\usepackage[all]{hypcap} 
\usepackage[utf8]{inputenc}
\usepackage[T1]{fontenc}
\usepackage{mathptmx}
\usepackage{etoolbox}

\draft 

\begin{document}


\title[Time Evolution of the Pinch Region of a Deflagration Plasma Accelerator]{Time Evolution of the Pinch Region of a Deflagration Plasma Accelerator} 

\author{A. A. T. Jibodu}
    \email{ajibodu@stanford.edu}
\author{J. D. Strickland}
    \email{jdstrickland@stanford.edu}
\author{M. A. Cappelli}
    \email{cap@stanford.edu}
\affiliation{Stanford Plasma Physics Laboratory, Stanford University, Stanford, CA 94305, US\\}

\date{\today}

\begin{abstract}
A spectroscopic study of a plasma deflagration accelerator was carried out to investigate the temporal evolution of plasma density within the pinch region. A half-meter imaging monochromator, paired with a fast (10 MHz) camera operating at 1 MHz, was used to collect broadened chord-integrated spectral lines from the pinch region of a plasma deflagration device. Specifically, images of the $H_{\beta}$ and the $H_{\alpha}$ lines were taken - with the $H_{\alpha}$ used to find the background continuum. Voigt fits of the Abel inverted H$_{\beta}$ emission lines allowed for determination of the radial profile of the number density in the pinch at intervals of 1 $\mu$s. This provided insight into the formation, growth, and decay of the pinch in both the deflagration and detonation modes of the accelerator. It was found that the maximum density for the deflagration increased from $\sim 10^{20}$ m$^{-3}$ 1.4 cm away from the core of the pinch to $\sim 10^{23}$ m$^{-3}$ at the core of the deflagration pinch. In contrast, the detonation pinch featured a broader zone of relatively constant density on the order of $\sim 10^{20}$ m$^{-3}$. Furthermore, an energy balance of the plasma in the deflagration pinch and downstream, as informed by prior work, was done to get an estimate of temperatures in the core of the pinch where measurements are not currently possible revealing potential temperatures on the 550 eV in the core of the pinch suggesting departures from some of the assumptions in the analysis. 

\end{abstract}


\maketitle 

\section{Introduction}
\label{Intro}

\begin{figure}
\centering
\includegraphics[scale=0.55]{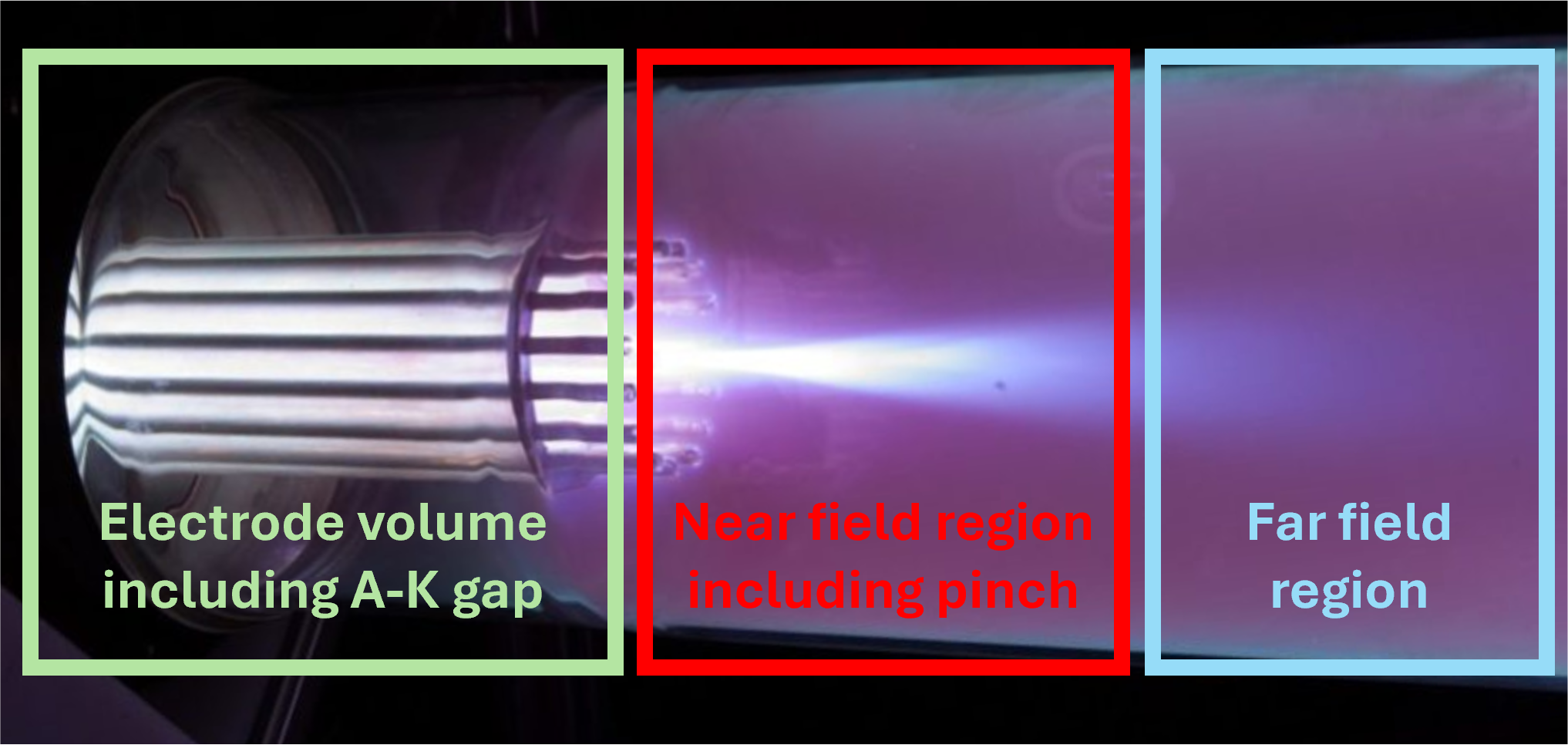}
\caption{Long exposure (5 s) image of the CHENG device during operation prominently showing the deflagration pinch region - the area of interest for this experiment.}
\label{LongExpImage}
\end{figure}

Pulsed coaxial plasma accelerators have found uses in various applications, including edge localized mode (ELM) studies \citep{garkusha2009latest}, fusion refueling \citep{WoodruffPRL, schaer1995coaxial, LeonardPRL}, and material processing studies \citep{liu1999investigation, kobayashi2001novel}, among others. Operation typically starts with gas in the anode-cathode (A-K) gap of a coaxial electrode set - or electrode volume - charged to a high potential difference, breaking down to generate a current-carrying plasma. This current-carrying plasma induces an azimuthal magnetic field which, in turn, results in a $\mathbf{J}\times \mathbf{B}$ Lorentz force, accelerating the plasma down the length of the electrodes. At the end of the electrodes, the current lines bow and cant to maintain contact with the electrodes as the plasma exits the barrel. This leads to a radial compressive component of the Lorentz force, resulting in a high-density pinch on the axis of the device. These steps are illustrated in Fig. \ref{LongExpImage} - a long exposure image of one such accelerator in operation with breakdown in the electrode volume, the \textit{near field} - in which dynamics are largely governed by the pinch - and the \textit{far field} in which the plasma is allowed to expand without strong influence of the pinch. Depending on the extent of gas filling of the A-K gap when the high voltage is introduced, these accelerators can be operated in two modes, referred to in the literature as the "deflagration" and "detonation" mode.\cite{cheng1970plasma, cheng1973experimental, loebner2015evidence, LoebnerBranching, cheng1971application, cheng1991deflagration, subhankar2024deflagration}

The deflagration mode is triggered when the high voltage is introduced prior to a significant gas fill of the A-K gap such that the current-carrying plasma, once generated, is allowed to accelerate into vacuum. This mode is characterized by a diffuse ionization zone in the A-K gap, longer pinch times, and higher exit velocity than the second mode and has been shown to have potential uses in lab-scale replication of astrophysical phenomena and propulsion.\cite{underwood2017plasma, cheng1971application, cheng1991deflagration, prathivadi2023characterization,subhankar2024deflagration} The second mode, the detonation (also referred to at times as the "snowplow" mode\cite{cheng1984deflagration}) is triggered when the high voltage is introduced after significant filling of the A-K gap such that the current carrying plasma is forced to accelerate into neutral gas, ionizing and sweeping it up in the process. This mode is characterized by a thinner shock-wave-like ionization zone as the plasma accelerates into the neutral gas, and a higher density and temperature pinch. Furthermore, in high-energy configurations of this mode, the $m = 0$ sausage instability (an axisymmetric instability mode in which local perturbations in a current-carrying plasma column induce alternating bulges and and constrictions with high magnetic fields, resembling a string of sausages) arises, producing populations of highly energetic electrons and ions. As such, this mode is conducive to nuclear fusion and neutron generation applications.\cite{shumlak2020z, stepanov2020flow, forbes2019progress} More thorough explanations of the two modes, their characteristics, and nuances can be found in the literature\cite{LoebnerBranching, Jibodu_Ballande_Cappelli_2023, Loebner}. Due to the application to fusion, the detonation mode has garnered interest over the years as it is the mode of operation of the well-studied dense plasma focus.\cite{krishnan2012dense,rawat2015dense, lerner2011theory}

The Stanford Co-axial High ENerGy (CHENG) device is one such coaxial plasma accelerator, capable of operating in the deflagration or detonation modes\cite{underwood2021dual}. The device features an array of outer anode rods arranged in a circular pattern with a 5 cm diameter and centered on a single central rod cathode. This was chosen instead of a solid outer electrode to allow for diagnostics of the A-K gap. with a clear polycarbonate sleeve around the outer anode rods providing gas containment. The system is typically powered by 56-$\mu$F capacitor banks capable of being charged to 20 kV. Upon charging the capacitors, which are connected to the electrodes via low-inductance co-axial transmission lines, gas is injected into the breech of the device using a custom-fabricated fast acting valve.\cite{loebner2015fast} This induces the deflagration process. The deflagration typically lasts about 10 $\mu$s after which, the generated magnetic field from the deflagration inductively charges the capacitors (the ring-down of the circuit-load system), leading to subsequent detonation mode breakdown events as the A-K gap is now pre-filled with neutral gas in the wake of the deflagration. This ring-down phenomenon and other characteristics of the CHENG have been discussed in prior publications\citep{LoebnerBranching, Jibodu_Ballande_Cappelli_2023, Loebner_Thesis}

Some prior studies of our generated plasma deflagration have focused on characterizing the plasma \textit{downstream} of the pinch, providing time and space-resolved measurements of plasma density, temperature, potential, and velocity \citep{Jibodu_Ballande_Cappelli_2023}. However, most measurements of the pinch region so far have been qualitative or have lacked time resolution \cite{Loebner_Thesis}. This current paper aims to provide time-resolved insight into the pinch formation and dynamics using optical emission spectroscopy (OES) - a passive diagnostic technique that uses characteristics of spectral lines from a plasma to gain insight into the plasma properties - typically, plasma density and temperature. While it is one of the more widely used plasma diagnostics, it is typically time-integrated when used with pulsed plasma accelerators like the CHENG device. We demonstrate that with a fast framing camera, this diagnostic may be time-resolved to provide deeper insight into the dynamics of the pinch and the produced compression. 

\section{Theoretical Background}
\label{TheoryBackground}

Optical emission spectroscopy concerning plasma involves utilizing the light emitted from plasma to understand its compositional elements. By dispersing the light into its component spectral lines and considering the characteristics of the individual lines - typically their line shape - insight can be gained into the temperature, density, presence of impurities, and a number of other characteristics of the plasma. For the purposes of this work, we consider two mechanisms that affect the line shape and lead to broadening of what would otherwise be narrow spectral lines upon dispersion. Stark broadening and Doppler broadening. Other mechanisms, including natural, quasi-static, and impact broadening, are excluded due to the insignificance of their effects when considered individually.

Stark broadening is largely due to plasma density. Typically, when bound electrons transition from one state to another, they absorb or emit light at specific wavelengths (or energies) equal to the amount of energy needed to make the transition. External electric fields would distort that process and affect the amount of energy needed for or given off by that transition. An electron being pulled away from the atom/molecule to which it is bound by an external electric field would require less energy to make a jump to a higher level, while giving off more energy when coming to a lower level. Conversely, an electron being pushed towards its nucleus would require more energy to jump to a higher level and give off less energy when settling to a lower level. In a plasma of sufficient density, any given electron would have its energy levels disrupted by the fluctuating electric fields of the surrounding ions and electrons. As such, any measured spectral line would be broadened. This broadening is known to be well approximated by a Cauchy Lorentz (Lorentzian) distribution \citep{GRIEM1976331}, the normalized version of which is given as
\begin{equation}
    L(x,f_L) = \frac{f_L}{2\pi\left[\left(f_L/2 \right)^2+(x-x_0)^2\right]}
\end{equation}
where $f_L$ is the full width at half maximum (FWHM) of the distribution and $x_0$ is the center of the distribution. The FWHM is a complex function of plasma density such that the density can often only be approximated using computation models and empirical tabulations \citep{Gigosos_2014, Aparicio_1998}. The hydrogen lines' FHWM, in $\text{\AA}$, have been approximated as 
\begin{equation}
    f_L = \left(2.50 \times 10^{-9}\right)\alpha_{1/2}N_e^{2/3}
\end{equation}
where $N_e$ refers to the electron density measured in cm$^{-3}$, while $\alpha_{1/2}$ denotes the half-width parameter. This parameter was selected with reference to the tabulated data provided by Grim, complemented by computer simulations. It is important to note that this half-width parameter varies depending on the spectral line under examination; however, it is systematically tabulated by \citet{Griem} specifically for the hydrogen lines pertinent to this research.

Doppler broadening, on the contrary, is largely due to the temperature of the plasma. As the atoms to which the electrons are bound move back and forth, the emitted spectral lines also shift to longer or shorter wavelengths. The velocity distribution in a plasma, if assumed to be Maxwellian, is Gaussian \citep{Griem_1997}. As such, the Doppler broadening is also a Gaussian with a normalized form of

\begin{equation}
    G(x,f_G) = \frac{2}{f_G}\sqrt{\frac{ln(2)}{\pi}}e^{-4ln(2)\left((x-x_0)/f_G\right)^2}
\end{equation}
where $f_G$ is the Gaussian FWHM. Unlike the relationship between the Lorentzian FWHM and the density, the Gaussian FWHM, in $\text{\AA}$, is a straightforward function of the temperature given as 

\begin{equation}
    f_G = (7.16 \times 10^{-7})\lambda\left(\frac{T}{M}\right)^{1/2}
\end{equation}
where $\lambda$ is the wavelength in $\text{\AA}$, T is the ion temperature in K, and M is the atomic weight in units of atomic mass.

In practice, any spectral line from a given plasma source would have components of both Doppler and Stark broadening. This combination presents as the convolution of a Gaussian and Lorentzian, referred to as a Voigt profile, with a normalized form given as: 

\begin{equation}
    V(x,f_G,f_L) = \int_{-\infty}^{\infty} G(u,f_G)*L(u-x,f_L) \,du
\end{equation}
Fitting a given spectral line to this Voigt profile would yield both the Lorentzian and the Gaussian FWHM, allowing for simultaneous computation of the ion temperature and density. Computationally, the exact Voigt profile is difficult to resolve, but a so-called pseudo-Voigt profile fitting may be performed instead \citep{Thompson_Pseudo}. This allows for the approximation of the Voigt using a weighted sum of a Lorentzian and a Gaussian profile as:

\begin{equation}
    V(x,f_G,f_L) \simeq \eta L(x,f_L)+(1-\eta)G(x,f_G)
\end{equation}
where $\eta$ is the weighting parameter and is a function of the two FWHM values. A class of schemes have been developed for prescribing $\eta$ but a fairly simple and implementable scheme with 1\% accuracy is \citep{Thompson_Pseudo,Ida_Extended_pseudo}

\begin{equation}
    \eta = 1.36603\left(\frac{f_L}{f}\right)-0.47719\left(\frac{f_L}{f}\right)^2+0.11116\left(\frac{f_L}{f}\right)^3
\end{equation}
where $f$ is the effective Voigt FWHM and is itself a function of the Lorentzian and Gaussian FWHM values given by

\begin{equation}
\begin{split}
    f = \bigl[&f_G^5+2.69269f_G^4f_L+2.42843f_G^3f_L^2+\\&4.47163f_G^2f_L^3+0.07842f_Gf_L^4+f_L^5 \bigr]^{1/5}
\end{split}
\end{equation}

\section{Experimental Set-Up} 
\label{ExpSetup}

Measurements were conducted at a distance of 2 cm from the terminal end of the central copper electrode of the device. An f/4 lens with a focal length of 40 cm was employed for the collection of light, which was subsequently concentrated onto the entrance slit of a 0.5 meter Spex 500M monochromator utilizing an identical lens. The entrance slit was configured to have a horizontal aperture of 75 µm and was fully expanded vertically to 2.54 cm. The monochromator was equipped with a 1200 groove/mm grating designed to disperse the light. This dispersed light was then captured using a Shimadzu HPV-X2 unintensified CMOS camera. Despite the Shimadzu's capability to operate at 10 MHz with an exposure time of 50 ns per frame, the experiments were conducted at a frequency of 1 MHz with a 500 ns exposure time. Shorter exposure times were not used due to the inadequacy of light for proper analysis. The camera was configured to capture 128 frames at 1 µs intervals upon activation.


\begin{figure}
\centering
\includegraphics[scale=0.6]{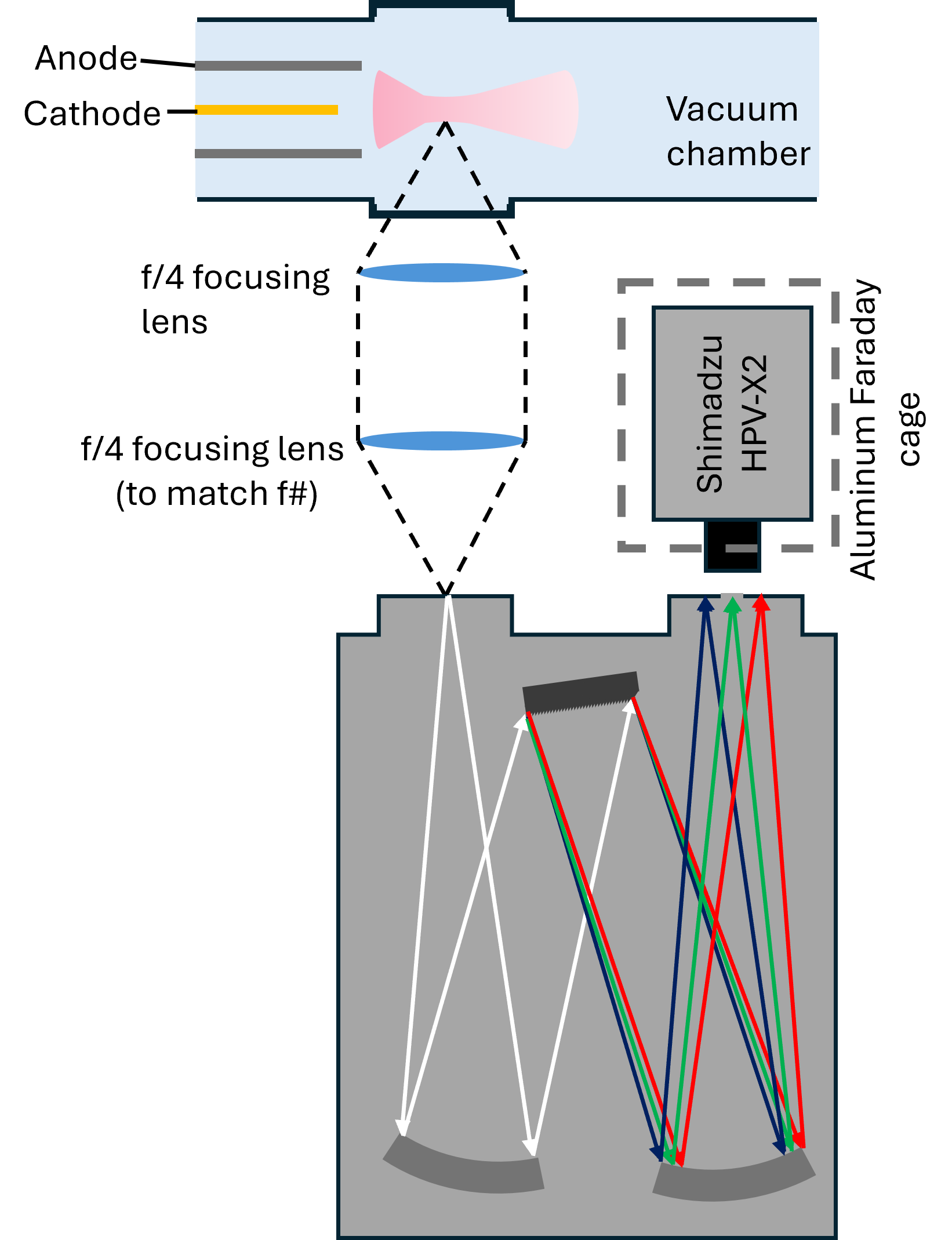}
\caption{The experimental setup included two f/4 lenses for focusing light on the entrance slit of the spectrometer half and a Shimadzu high speed camera gate for 1 $\mu$s with an exposure time of 500 ns}.
\label{Expsetupdiag}
\end{figure}

The Shimadzu camera was paired with a Nikon Micro-Nikkor 55 mm f2.8 lens, which had its iris fully open to collect as much light as possible. The exit slit assembly of the monochromator was removed to maximize the field of view. This allowed for an effective viewing region that spanned about 190 $\text{\AA}$ on the camera. \textbf{Lets put the resolution here?} A matte black tube was placed between the lens and the exit of the monochromator to prevent stray light - a black towel was placed around the setup to further prevent stray light. Discharges were generated on the CHENG with the monochromator scanning the region from 416 to 541 nm and from 636 to 676 nm in 8 nm steps to facilitate identification of constituents in the plasma and to establish a background. The region around $H_{\alpha}$ from 636 to 676 nm was found to be the most useful for establishing the background, as impurities affected most other regions, and $H_{\beta}$ was found to be both bright enough and broad enough for analysis with the camera's resolution. A flat continuum was assumed for this analysis, which was expected to be adequate considering the resolution at such low light levels. Five discharges were taken at each step and each was generated at a driving capacitor charge voltage of 5 kV, which corresponds to 700 J, with argon as the working gas. Hydrogen, as an impurity, was introduced into the system via ablation of the insulator at the breech of the device. The plenum pressure for the fast-acting valve on the CHENG device was maintained at a gauge of 40 psi (276 kPa). The valve charge voltage was maintained at 900 V \citep{Jibodu_Ballande_Cappelli_2023,loebner2015fast}. 

The camera was synchronized with the discharges using a Pearson current monitor (Pearson Electronics model 4997), which was used to monitor the discharge current in the capacitors for the CHENG. The oscilloscope was set to trigger when the current into the CHENG was at least 24 kA - signifying current flow while also above the noise limit for the oscilloscope, which would lead to erroneous triggering. The oscilloscope generated a negative TTL (Transistor-Transistor Logic) pulse, a standard digital signal used for synchronization, that triggered the camera to initiate recording while it was in its external standby state. The first 2 frames for each recorded frame were discarded as they had anomalous bright spots contaminating the data. It was hypothesized that this phenomenon resulted from the operational characteristics of the camera, specifically involving the accumulation of charges that were subsequently released onto the initial captured frames. Such frames were consistently observed when the system was triggered from the external standby mode, potentially attributable to the CMOS releasing charge accumulated during the waiting period for a trigger signal. The external trigger mode of the camera was avoided as it had more such frames. In addition to the issue with the first two frames, the camera had a built-in delay of 890 ns such that the first usable frame was 2890 ns from the triggered time. 

\begin{figure}
\centering
\includegraphics[scale=0.6]{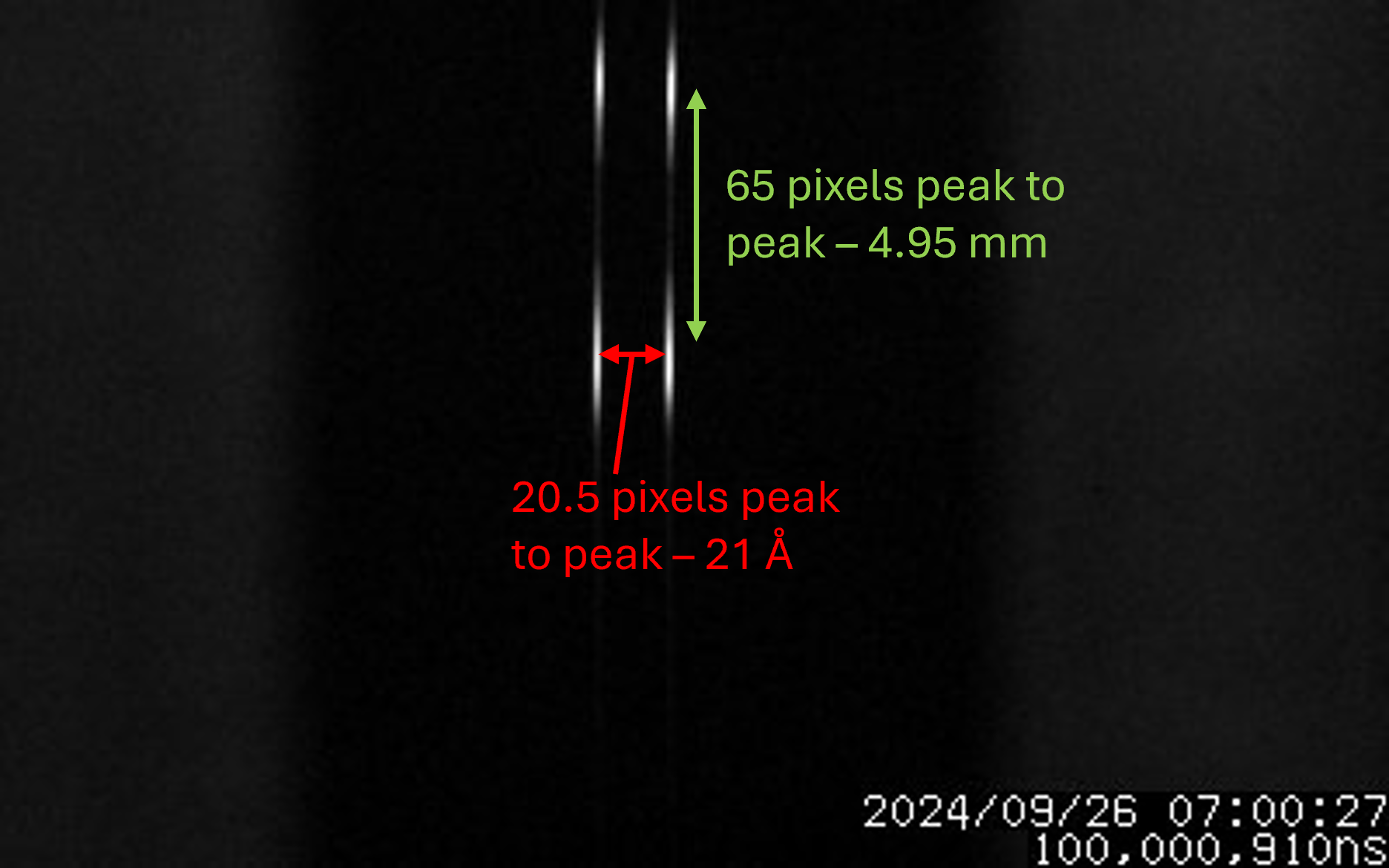}
\caption{A mercury lamp was used for spatial and spectral calibration. A spatial calibration factor of 0.076 mm/pixel was found with a spectral calibration factor of 1.025 $\text{\AA}$/pixel}
\label{MercuryCal}
\end{figure}

Focusing and spatial and spectral calibration were performed by placing a mercury lamp at the expected location of the pinch of the plasma jet and taking images of the spectral lines of the mercury doublet at 5769.6 $\text{\AA}$ and 5790.65 $\text{\AA}$ as shown in Fig. \ref{MercuryCal} in a manner similar to that prescribed by \cite{Loebner_Thesis}. This yielded a wavelength calibration factor of 1.025 $\text{\AA}$/pixel and a vertical length calibration factor of 0.076 mm/pixel. The wavelength calibration factor was further confirmed when scanning the spectrum, as a wavelength shift of 80 $\text{\AA}$ resulted in a shift of 78 pixels, reducing to 1.025 $\text{\AA}$/pixel. 

Flat-field and dark-field images were taken to correct for the recorded frames. The dark-field images were taken by closing the entrance slit of the monochromator. The flat field was taken by placing a piece of white paper, which was back-illuminated by fluorescent lights, at about the same focal distance as the spectral lines. 15 recordings of 128 frames each were taken and averaged out to get the flat-field. The paper was moved slightly with each recording so that its structure - visible in the recordings - would get averaged out. The general shape of the flat field was found to be consistent with pixel values generally decreasing from left to right and from top to bottom, such that the highest pixel values were generally in the top left and the lowest in the bottom right of the recorded frames. However, the percent drop was found to be sensitive to the amount of light coming into the camera such that at low light levels (close to dark field), the lowest pixel values were about 50\% of the highest pixel values, but the relative percent difference decreased to less than 5\% for high light input (close to saturation). The aperture of the lens was set to capture a light amount comparable to that produced by the $H_\beta$ line during the experiments.

Finally, the instrument broadening was characterized by directing the emission from a spectral lamp through the entrance slit. The resultant instrument broadening FWHM was about 2 $\text{\AA}$. This is typically modeled as Gaussian in nature, but presented as triangular because of the finite resolution of the camera.

In practical applications, any captured image of the spectral lines represents the chord-integrated emission of the plasma. Consequently, any direct analysis performed on this image will inherently be chord-integrated as well. Although this approach has intrinsic interest, the radial profile of the light emission is of greater utility, as it facilitates the determination of the radial profiles of plasma density and temperature. Conversion of chord-integrated emission data to radial emission profiles is possible at each wavelength using an Abel-inversion. This describes the radial profile, $f(r)$, as a function of a chord-integrated profile $F(y)$ as

\begin{equation}
    f(r) = -\frac{1}{\pi}\int_{r}^{\infty} \frac{dF(y)}{dy} \frac{1}{\sqrt{y^2-r^2}} \,dy
\end{equation}

This may be evaluated using a variety of different methods, as will be discussed in further detail in Section \ref{AnalysisandResults}. Implicit in the Abel inversion process is the assumption that the plasma is optically thin, meaning that reabsorption of light traveling in the plasma is minimal. In addition to being a requirement for Abel-Inversion, having a plasma that is not optically thin would result in artificial reduction of the line peak and increased broadening \citep{Rezaei16}. This optical depth requirement is met when $\kappa l<1$ where $\kappa$ and $l$ are the plasma absorption coefficient and the geometric thickness of the plasma. For the pinch region, $l$ can be taken to be 1 cm - the thickness of the region of highest density, although it is expected that there is a lower density plasma surrounding this dense pinch region\citep{Loebner}. $\kappa$ can be found as \citep{Cappelli_Thesis}
\begin{equation}
    \kappa = \frac{N_n\lambda^2A_{nm}f_{nm}(\nu)}{8\pi}e^{\bigl(\frac{hc}{\lambda k_bT_e}-1\bigr)}
\label{AbelInversionEq}
\end{equation}
where $N_n$ is the density of atoms in the higher energy level for the associated transition for the spectral line, $c$ is the speed of light, $\lambda$ is the wavelength of the transition, $A_{nm}$ is the Einstein spontaneous emission coefficient, $f_{nm}(\nu)$ is the line shape of the spectral line wavelength but evaluated at the associated frequency, $\nu$, for the spectral line. For the purposes of analysis, a Lorentzian line shape may be assumed, as that would be expected to be dominant in such high-density conditions. Evaluating at the center wavelength of the spread yields 
\begin{equation}
    f_{nm}(\nu) = \|\frac{2}{\pi d\nu}\|
\end{equation}
where $d\nu$ denotes the equivalent alteration in frequency, corresponding to the frequency representation of the change in wavelength $d\lambda$, which can be assumed to be $f_L$. In order to compute $d\nu$, it is necessary to consider $c = \lambda\nu$, where $\nu$ represents the frequency of light. This yields the relationship between changes in frequency and wavelength as
\begin{equation}
    d\nu = \frac{-c}{\lambda^2}d\lambda
\end{equation}
which allows for the evaluation of $f_{nm}(\nu)$. Focusing on the $H_\beta$ line and assuming a nominal peak plasma temperature of $\sim$ 50 eV based on previous work \citep{Loebner_Thesis}, a nominal $f_L$ of $\sim$ 40 $\text{\AA}$, an Einstein spontaneous emission coefficient of $.2 \times 10^8 s^{-1}$, and the pinch diameter of 1 cm, the optically thin assumption is valid when $N_n < 1.2 \times 10^{22} m^{-3}$. It should be stressed that this is the density of the upper energy state of the transition. Furthermore, hydrogen is an impurity in this experiment such that the plasma density may itself be higher than this limit without violating the optically thin condition. 

\section{Analysis and Results}
\label{AnalysisandResults}

\begin{figure}
\centering
\includegraphics[scale=0.485]{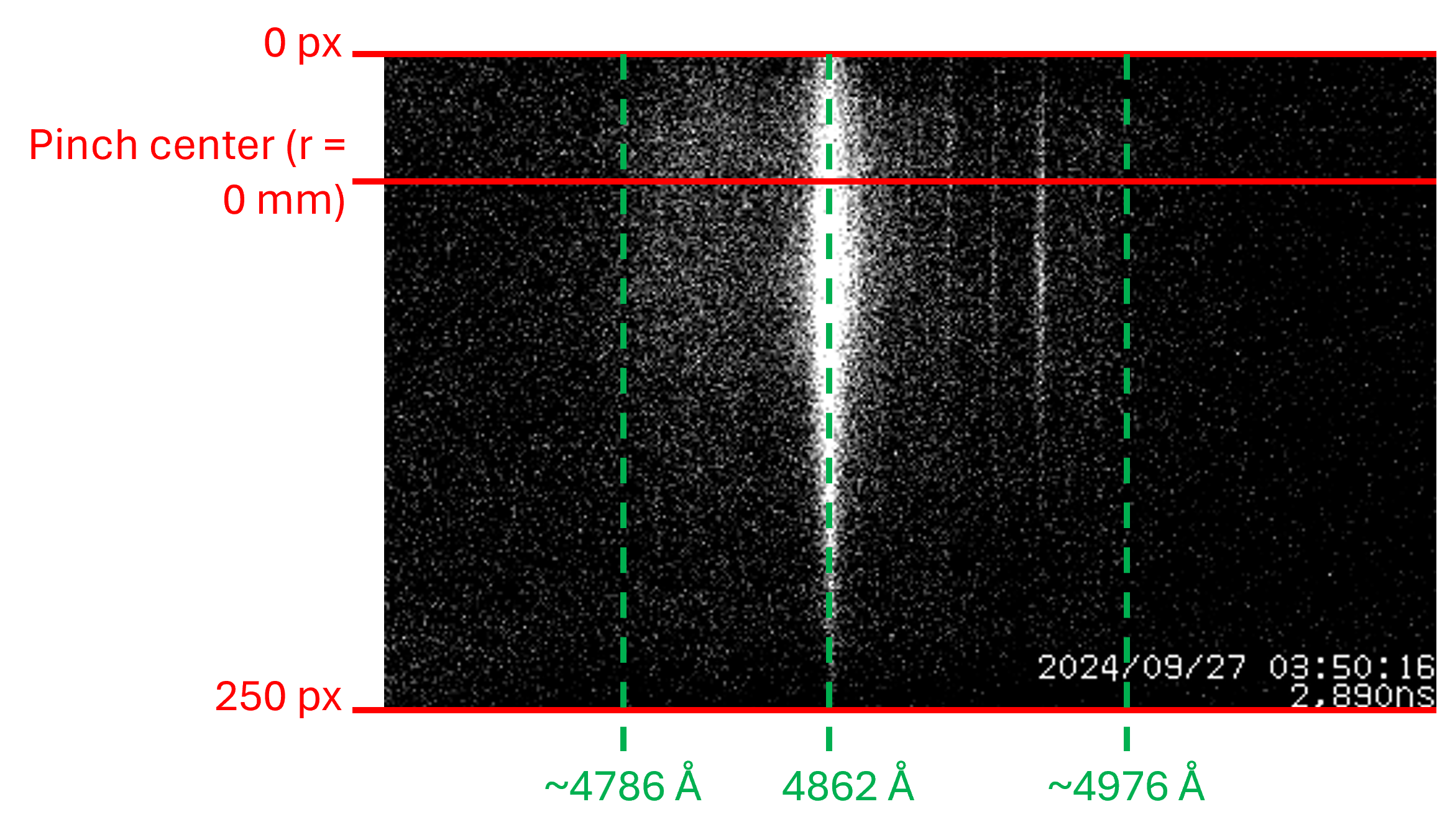}
\caption{Example frame featuring the broad $H_\beta$ line and some other spectral lines in the pinch region during a deflagration. The picture image spans 190 $\text{\AA}$ of the spectrum (area demarcated in green), and the center of the pinch is about 30 pixels from the top of the image} 
\label{HBetaframe}
\end{figure}

\begin{figure}
\centering
\includegraphics[scale=0.64]{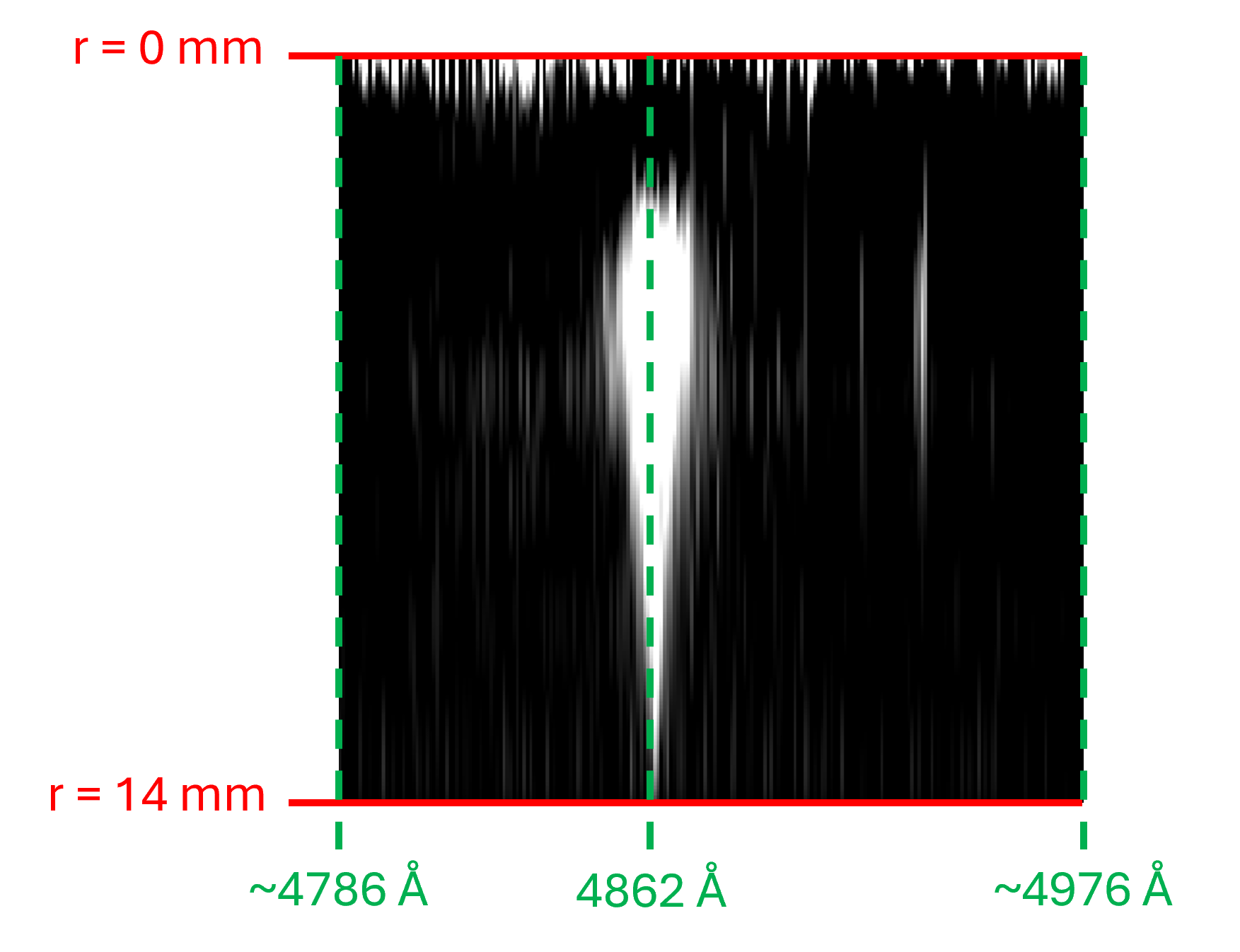}
\caption{Abel inverted image of an acquired emission image after dark-field and flat-field corrections. The image has been cropped to show the field of view and re-leveled for visibility}. 
\label{AbelInverted}
\end{figure}

Figure \ref{HBetaframe} shows an example of a raw data frame acquired during the deflagration process. This, and all images acquired for the $H_\beta$ lines, were Abel-inverted to obtain the radial emission profile per equation. \ref{AbelInversionEq}. MATLAB was used to generate a 5-segment cubic spline fit for each vertical slice, which was then numerically differentiated, and the integral was solved to obtain the radial emission profile at that slice. The center (r = 0) was taken to be y = 30 pixels, where y = 0 pixels is the top of the raw image. This formulation yielded radial emission patterns similar to those shown in Fig. \ref{AbelInverted}. 

\begin{figure}
\centering
\includegraphics[scale=0.41]{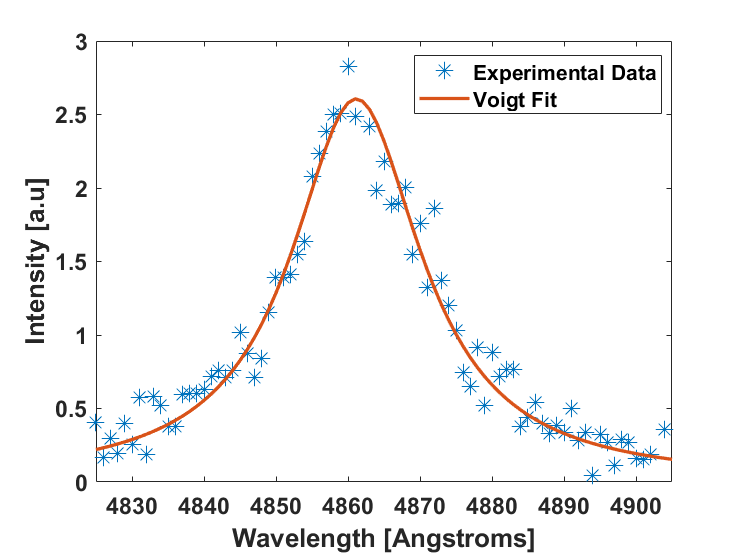}
\caption{Sample Voigt fit to $H_\beta$ data with a calculated Lorentzian FWHM of 22.0 $\text{\AA}$ and no Gaussian component. This was taken at r = 6.25 mm at t = 3$\mu$s}. 
\label{VoigtFit}
\end{figure}

A MATLAB code was also used to go row by row to perform the Voigt fit, from which the Lorentzian and Gaussian FWHMs were extracted. The fits - an example of which is shown in Fig. \ref{VoigtFit} - were found to be largely Lorentzian in nature, coming from the outermost layer of the plasma until about 4 mm from the core of the plasma at which point the fits diverged, producing nonphysically large Gaussian components on the order of 10s of $\text{\AA}$. 

It should be noted that a 50 eV plasma is expected to produce a Gaussian FWHM of about 2.5 $\text{\AA}$. Thus, the non-physically large Gaussian components found may be due to error accumulation that is known to occur during the Abel inversion process, where the error increases closer to r = 0 \citep{Ramsey_Abel_error}. Furthermore, as discussed, the pixel-pixel variation for which the flat-field correction was done was dependent on the light level. Closer to r = 0, the difference in light levels from the brightest spot to the noise background was greater, resulting in increased error due to the pixel-pixel variation. Typical calculated Lorentzian, Gaussian, and effective FWHM are illustrated below in Fig. \ref{DifferingFWHM}.

\begin{figure}
\centering
\includegraphics[scale=0.45]{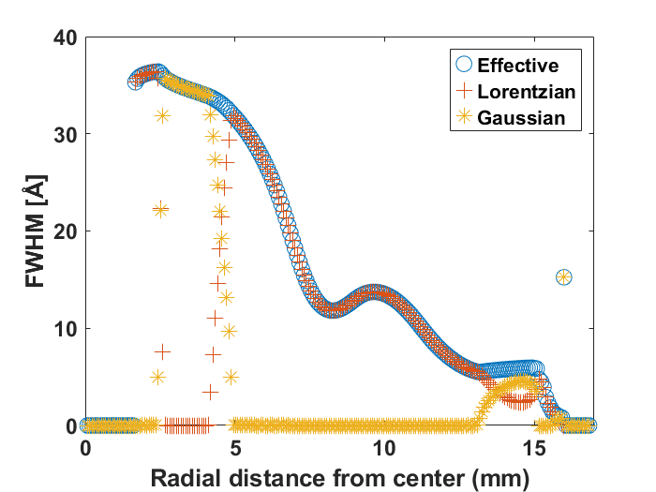}
\caption{Effective Voigt FWHM compared to the Lorentzian and Gaussian components for a typical $H_{\beta}$ image. The fits are largely Lorentzian in nature but show nonphysically high Gaussian components close to the core of the pinch, suggestive of externalities - numerical, physical, or both. There is a burn-out of neutral hydrogen about 25 pixels, which corresponds to about 2 mm}. 
\label{DifferingFWHM}
\end{figure}

\begin{figure}
\centering
\includegraphics[scale=0.45]{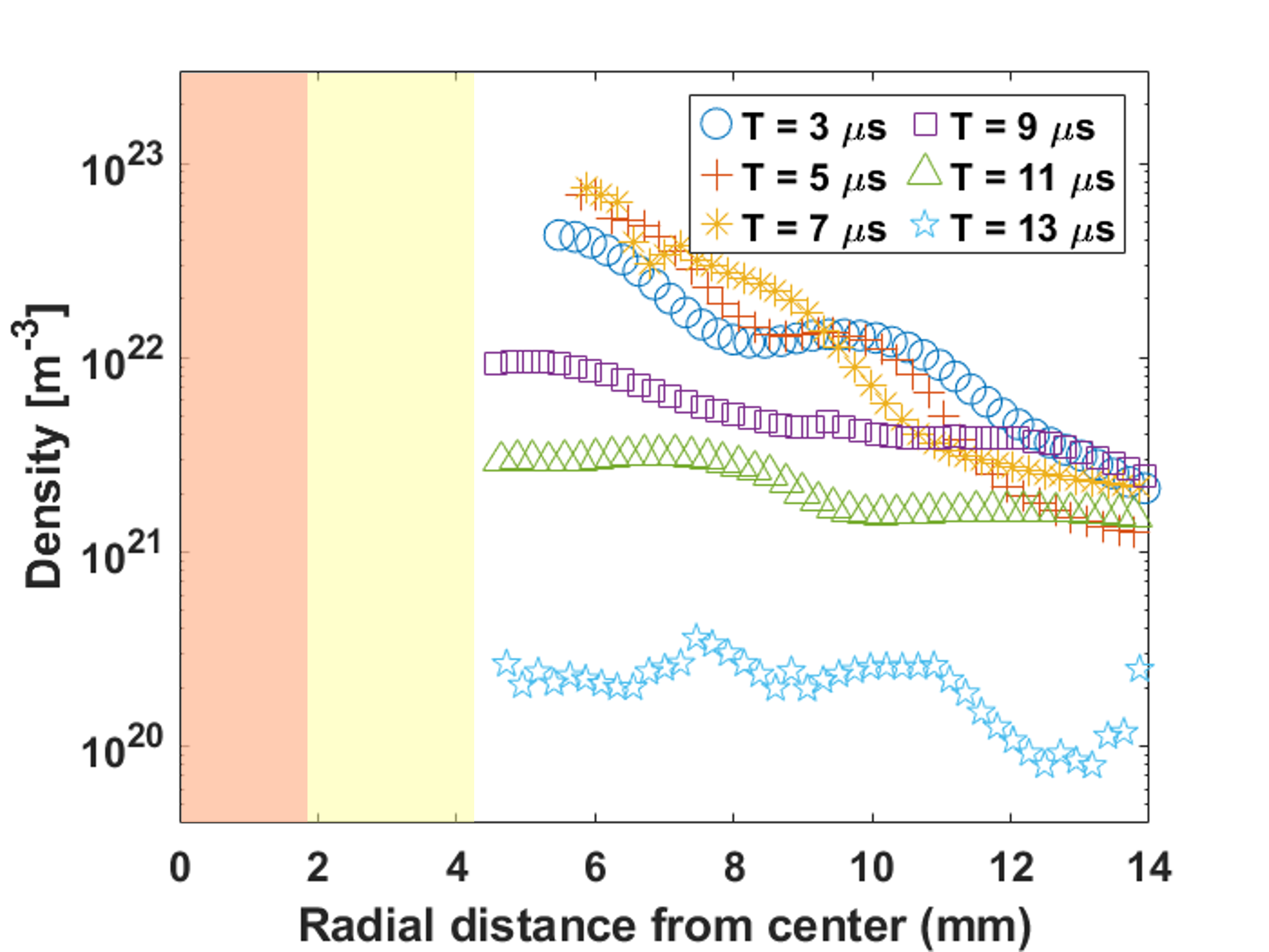}
\caption{Time-resolved calculated plasma densities from $H_{\beta}$ line Voigts fits as found after averaging across 5 argon deflagration discharges at 5 kV. The yellow region represents regions with no Lorentzian components but nonphysically large Gaussian components. The red region denotes the burn-out region in which there was no $H_{\beta}$ data, suggesting a lack of neutral hydrogen.} 
\label{DefTimeZones}
\end{figure}

Going even closer to the core of the pinch - about 2 mm away - there was a region with no light or data suggestive of a "burnout" in which no quantifiable neutral hydrogen was present and all hydrogen was ionized. This was previously seen in \citet{Loebner_Thesis} with time-integrated $H_\alpha$ measurements in the pinch. Ignoring the region of large Gaussian components and the burn-out region, the plasma density was found to increase with proximity to the pinch core as shown in Fig. \ref{DefTimeZones}. This matches the expected trends from previous work. As shown, the calculated densities were found to be the same during the duration of the pinch up to about 8 $\mu$ s, which corresponds to the expected time frame of the pinch.   

Note that by 9 $\mu$s, the density had started to drop. By 11 $\mu$s, the pinch was largely decayed, resulting in a drop in densities. The calculated peak densities using the Lorentzian FWHM measured were 7 $ \pm 3 \times 10^{22}$ m$^{-3}$. Assuming a Gaussian fall off, the projected peak density in the core was expected to be 4 $ \times 10^{23}$ m$^{-3}$, which matched the value of the prior work. The prior work also used Schlieren imaging to determine the general shape of the radial density profile in the deflagration pinch region. Considering the times during the pinch's lifetime (prior to 9 $\mu$s) and the location of the measured - less than one accelerator diameter away - there was an expected 6x density increase from about 6 mm to the peak core density based on the Schlieren maps, which matches the expected core density calculated assuming a Gaussian fall off.

\begin{figure}
\centering
\includegraphics[scale=0.45]{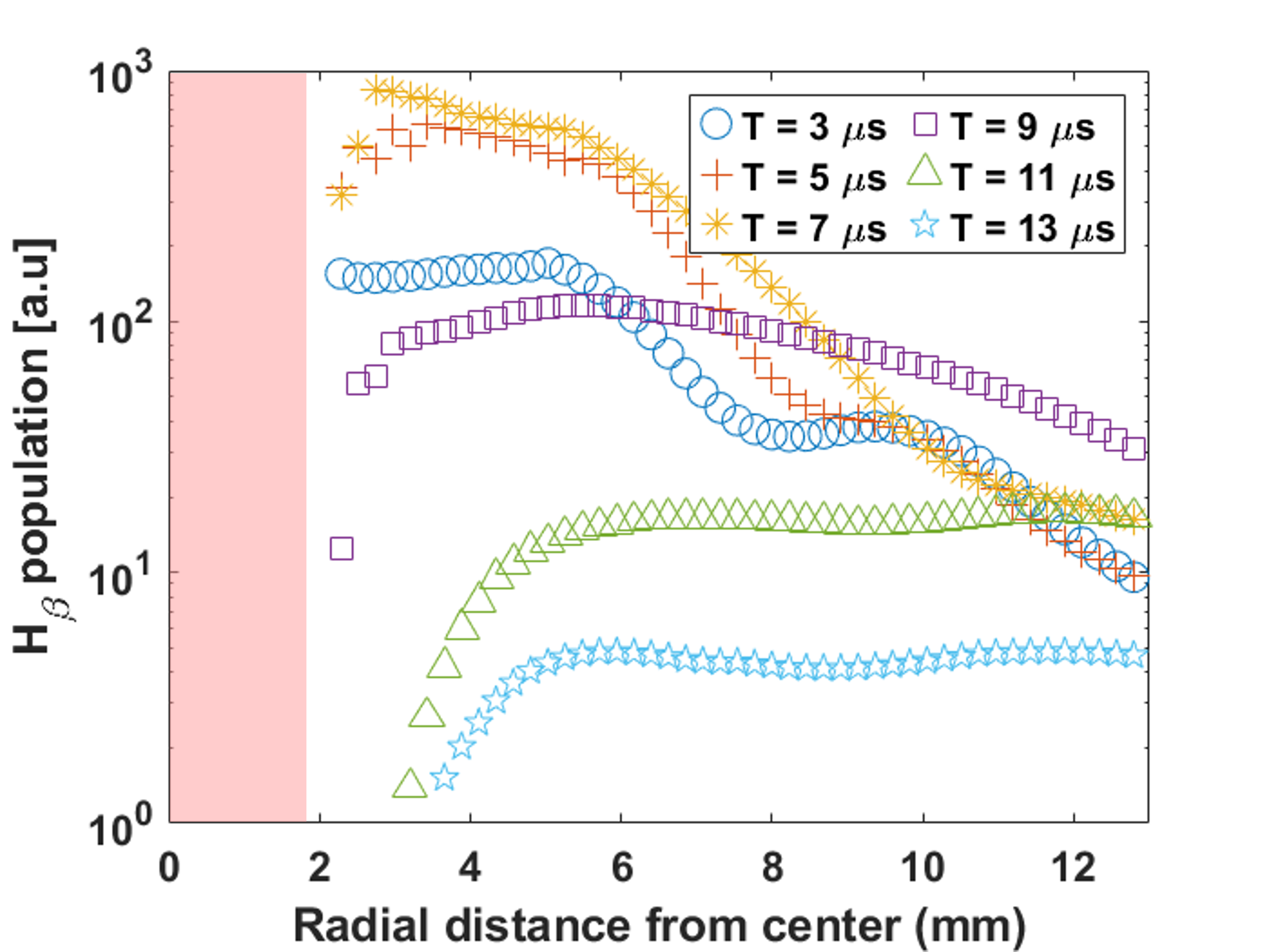}
\caption{Time-resolved $H_{\beta}$ population found by integrating the area under the curves of the $H_{\beta}$ emission for the deflagration. Note the general increase in the population values during the early pinch times. Also, note the peak near the burn-out region during the lifetime of the pinch  with a gradual drop and outward movement of the $H_{\beta}$ population as the pinch decays. This was on 5 kV argon shots with hydrogen as an impurity}. 
\label{DefPopDensZones}
\end{figure}

The $H_{\beta}$ burn-out region was further quantified considering the area under the curve of the inverted image that provides a measure of the radial profile of the density of the population. As shown in Fig. \ref{DefPopDensZones}, the $H_{\beta}$ population increased over time until a maximum was reached around 6 $\mu$ s, then decreased as the pinch decayed. Of note was the oscillatory behavior at early pinch times in both the density curves, but especially in the population densities, suggestive of the kink instability. This has been known to appear in such devices, has been seen previously in video recordings and Schlieren imaging of this device, and was discussed at length by \citet{underwood2019hydromagnetic}. It should also be noted that during the pinch, the $H_{\beta}$ population is maximum around the burnout region. Intuitively, this follows as the energy necessary for the transition would be readily supplied by the pinch. The drift of the bulk of the $H_{\beta}$ population from the center at later times is likely due to the expansion of the neutral hydrogen from the high pressure pinch core with the decay of the magnetic fields compressing the pinch.

\begin{figure}
\centering
\includegraphics[scale=0.45]{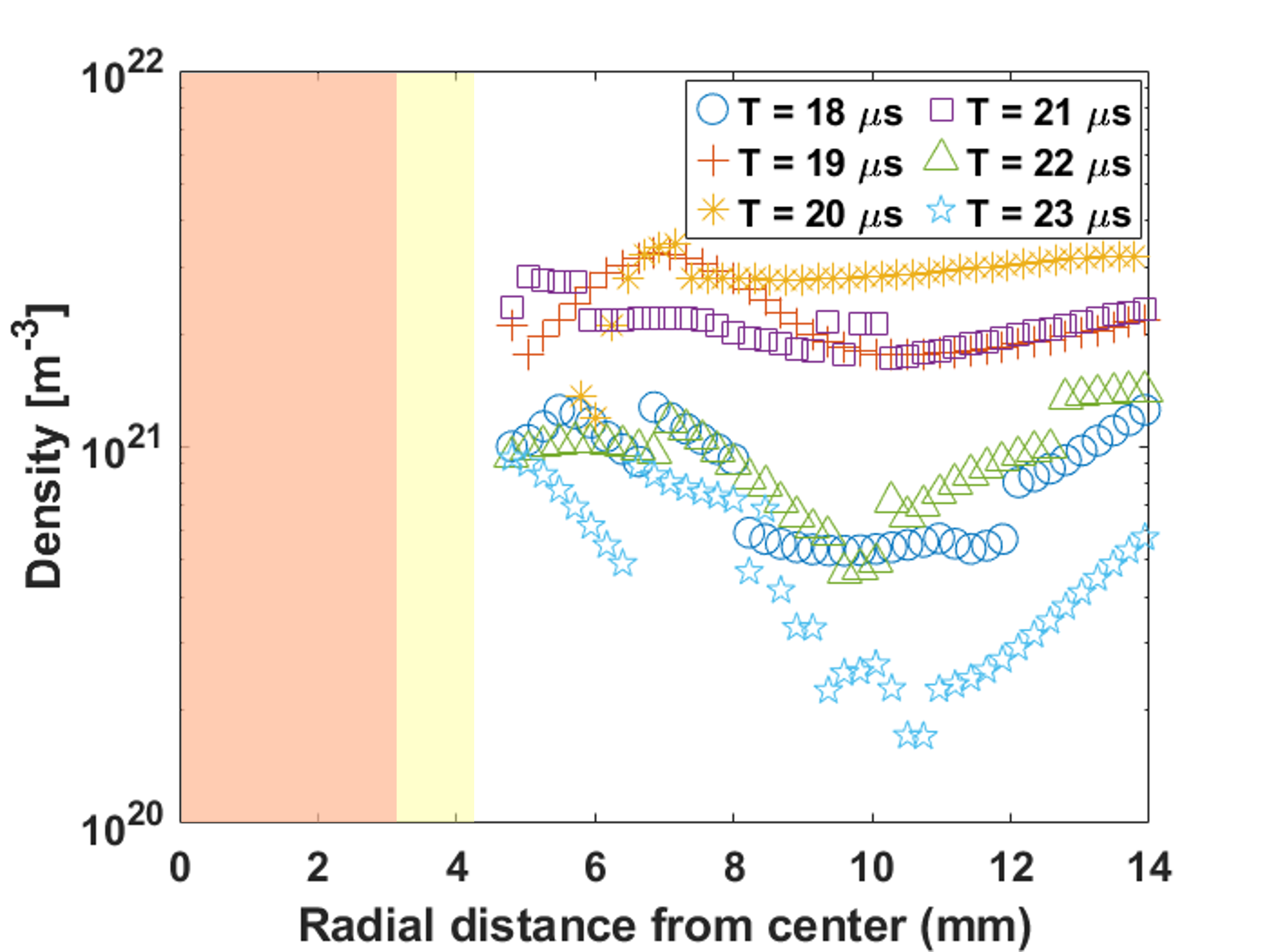}
\caption{Time-resolved calculated plasma densities from $H_{\beta}$ line Voigt fits for the first subsequent detonation discharge during an argon 5 kV experimental run. The yellow region represents regions with no Lorentzian components but nonphysically large Gaussian components. The red region denotes the burn-out region in which there was no $H_{\beta}$ data, suggesting a lack of neutral hydrogen. Note that the traces shown are in time steps of $\mu$s because of the much shorter pinch time}. 
\label{DetTimeZones}
\end{figure}

\begin{figure}
\centering
\includegraphics[scale=0.45]{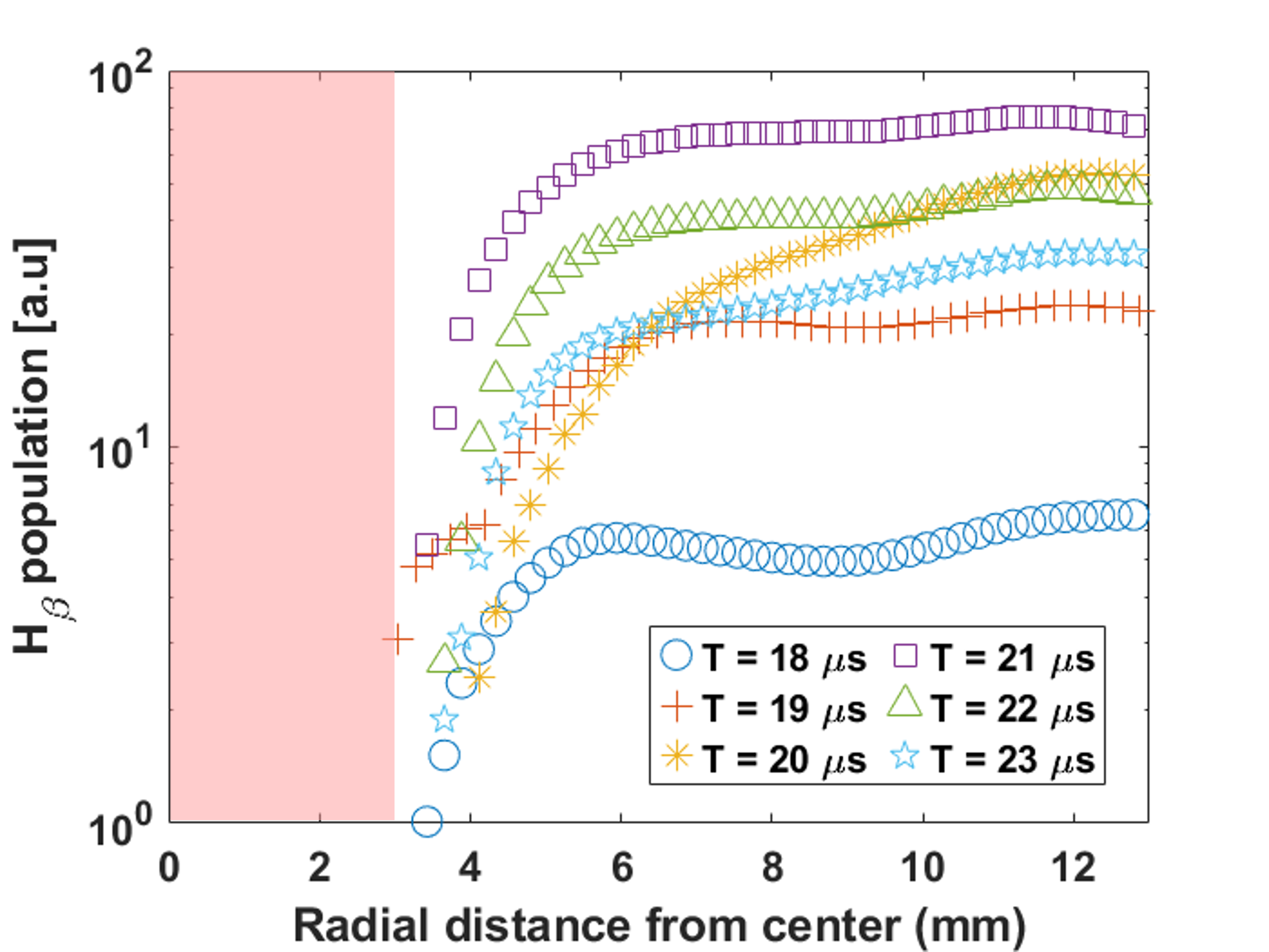}
\caption{Time-resolved $H_{\beta}$ population found by integrating the area under the curves of the $H_{\beta}$ emission for the first subsequent detonation. Note the general flatness of the population density across the region aside from the very core pinch region. Also note the faster rise and drop of the population during the detonation relative to the deflagration. This was on 5 kV argon shots with hydrogen as an impurity}. 
\label{DetPopDensZones}
\end{figure}

For the first subsequent detonation, as shown in Fig. \ref{DetTimeZones}, the plasma in the pinch region was present for a much shorter time period, resulting in a faster rise and fall of the measured densities. As discussed earlier, the detonation mode presents itself as more of a shock wave than a long-lasting plasma jet.  Furthermore, this detonation is a relatively lower energy discharge - relying on energy left in the system after the deflagration. As such, the densities generated are at least an order of magnitude less - reaching a peak measured density of 3 $ \pm 1 \times 10^{21}$ m$^{-3}$. This follows intuition when considering that there is little evidence of the detonations in Fig. \ref{LongExpImage}, which were integrated over the entire ring-down period of that discharge. In addition to the lower density, the lower energy results in a weaker pinch effect, which, in combination with the shockwave-like nature of the detonation, resulted in relatively flat contours of the calculated densities. Unlike regions of largely Lorentzian fits, then Gaussian fits closer to the core in the deflagration discharge, then the burn-out, the Voigt fits of the $H_{\beta}$ lines during the detonation produced Lorentzian fits until the burn-out region. Interestingly, the burn-out region of the detonation is larger than in the deflagration. This is further seen in the $H_{\beta}$ population as shown in Fig. \ref{DetPopDensZones}.

As illustrated in the figure,  the population of the excited $H_{\beta}$ during the detonation is less than the population of the deflagration. This is as expected and follows the density of the plasma in general. The extent of the burn-out region is counterintuitive as the lower energy going into the discharge and, subsequently, the pinch should result in lower temperatures in the core of the pinch, allowing for a smaller zone of complete ionization. However, what is not immediately apparent is that the lower density in the pinch and in the core would also result in higher ionization. As such, the probability of $H_{\beta}$ emission is a combination of both the temperature and the density of the pinch core - both of which would require further work to get insight into. Unlike Schlieren for deflagration, there is no prior comparison of the outer plasma to the core plasma densities, which may be similarly leveraged.

Calculating the temperatures for both the deflagration and detonation with the acquired data proved to be difficult. As discussed, the fits were largely Lorentzian with little to no Gaussian components, and when there were Gaussian components, they were often non-physically large. That said, the insight into the density provides a method for finding an upper limit to the temperature in the pinch core region for the deflagration using data in the far field of the device. According to \citet{PiriaeiQLP}, the time-integrated energy flux, as referred to in $E$, can be and has been determined utilizing a quadruple Langmuir probe. Conservation of energy may then be invoked, assuming little to no loss during the expansion from the pinch region to the QLP locations, to find this temperature upper limit. Experiments with argon have yielded $E$ values on the order of 22 kJ/m$^2$, which,  assuming a uniform jet front, allows the equation of energy balance to be written as

\begin{equation}
     \frac{3}{2}(n_ek_bT_e+n_ik_bT_i)V_{pinch} = EA_{flow}
     \label{energyconseq1}
\end{equation}
where $A_{flow}$ is the cross-sectional area of the flow at the QLP location ($\pi r^2$ where r = 6 cm is the expected radius of the plasma jet based on the tube around the CHENG) and $V_{pinch}$ is the volume swept out by the particles in the high density pinch region. A conservative estimate for $V_{pinch}$ may be found by taking the velocity in the pinch region to be half the velocity upon expansion (which may itself be taken to be nominally 30 km/s per prior work), a pinch lifetime of 8 $\mu$s - about when the pinch density starts dropping as discussed -  and the core of the pinch to be 1 cm in diameter per Schlieren imaging from previous work. Furthermore, assuming a nominal density of 1 $\times 10^{23}$ m$^{-3}$ for this region, quasineutrality (i.e. $n_e$ = $n_i$ = $n$) and local thermal equilibrium (i.e. $T_e$ = $T_i$ = $T$), the upper limit $T$ can be found to be $\sim$ 550 eV. Even with this upper limit temperature, which assumes that all the energy in the far-field goes through the pinch along with a number of very conservative assumptions, the Doppler FWHM would be $\sim$ 9 $\text{\AA}$ - still much smaller than the Gaussian FWHM found by Voigt fitting.

Admittedly, with the more realistic 50 eV, the expected Gaussian component would be on the order of the instrument broadening, suggesting that it may be impossible to calculate the temperature using this current experimental setup. This may be avoided by using a larger spectrometer, which would provide more resolution, albeit at the cost of light. The drop in light may be countered using an optical intensifier. There are plans to acquire a Lambert HiCATT intensifier to use in combination with the Shidmazu camera on a full-meter double monochromator already in-house.

Of note and interest in both deflagration density measurements were oscillations suggestive of the kink instability known to appear in such devices, seen previously in videography and Schlieren imaging of this device, and discussed at length by \citet{underwood2019hydromagnetic}. They seem especially prominent in the deflagration at earlier times, as expected, and are on length scales comparable to those seen in the Schlieren imaging. However, more work and higher resolution are needed to validate this as the same phenomenon. 

\section{Conclusions}
\label{Conclusions}

In conclusion, this study of the CHENG plasma deflagration accelerator provides valuable insights into the dynamic behavior of the pinch of a deflagration accelerator. Time-resolved optical emission spectroscopy - made possible by focusing a high-speed camera on the exit slit of a monochromator -  was performed on the pinch region of the accelerator. The experimental results allowed for a greater understanding of the lifetime of the pinch, calculation of an upper limit of the temperature of the pinch, and knowledge of the time variation of the densities within the pinch. Direct temperature measurements were not possible because of resolution and instrumental limitations. Density measurements close to the pinch core were limited because of numerical anomalies and neutral hydrogen burn-out, but leveraging fitting models and prior Schlieren work allowed for estimation of densities in the core using the calculable densities away from the core region. Analysis corroborated prior time-integrated optical emission studies and Schlieren imaging results. An energy balance calculation allowed one to determine an upper temperature limit in the core of the pinch. Further work is needed to actually calculate time-resolved temperatures within the pinch region. In particular, a higher resolution is needed, which might be accessible with a large monochromator and an intensified camera. 

\section{Funding}
This work is supported in part by a Lawrence Livermore National Laboratory DSTI Fellowship (grant number B653347); the Department of Energy (grant number DE-SC0021255); and the Air Force Office of Scientific Research (grant number FA9550-21-1-0016)

\section{Declaration of Interests}
The authors report no conflict of interest.

\section{Data Availability Statement}
The data that support the findings of this study are available from the corresponding author upon reasonable request.

\bibliography{TimeResolvedBib}

\end{document}